# Low Frequency Dielectric Loss of Metal/Insulator/Organic Semiconductor Junctions in Ambient Conditions

*R. Ledru[a], S. Pleutin[a,b]\*, B. Grouiez[a], D. Zander[a], H. Bejbouji[a,b], K. Lmimouni[b], and D. Vuillaume[b]*

[a] CReSTIC, University of Reims, Moulin de la Housse
F-51687 Reims Cedex 2 (France).

[b] Institute for Electronics, Microelectronics and Nanotechnology (IEMN), CNRS, University of Lille, BP 60069, Avenue Poincaré,
F-59652 Cedex, Villeneuve d'Ascq (France).

\* Corresponding author, E-mail: stephane.pleutin@isen.iemn.univ-lille1.fr

Keywords: Metal-Oxide-Organic semiconductor junctions, dielectric relaxation, anomalous diffusion, admittance spectroscopy

Abstract: The complex admittance of metal/oxide/pentacene thin film junctions is investigated under ambient conditions. At low frequencies, a contribution attributed to proton diffusion through the oxide is seen. This diffusion is shown to be anomalous and is believed to be also at the origin of the bias stress effect observed in organic field effect transistors. At higher frequencies, two dipolar contributions are evidenced, attributed to defects located one at the organic/oxide interface or within the organic, and the other in the bulk of the oxide. These two dipolar responses show different dynamic properties that manifest themselves in the admittance in the form of a Debye contribution for the defects located in the oxide, and of a Cole-Cole contribution for the defects related to the organic.

# 1. Introduction.

As Organic Field-Effect Transistors (OFET) are now close to applications questions about their reliability under realistic atmospheric conditions become more and more important.[1] Electrical instabilities are evidenced, for instance, in the bias stress effect where under a prolonged application of a gate potential the device characteristics, such as the threshold voltage, evolve with time.[1,2] Several mechanisms were proposed to understand these degradations of the device properties. Some are intrinsic mechanisms such as trapping of charges, in the organic film[3-5] or at the organic/oxide interface,[6] or pairing of mobile charges to heavier bipolarons.[7] Some others are extrinsic mechanisms needed especially when the measurements are done in the air since humidity was shown to amplify these instabilities.[8] An example of such mechanism was proposed recently based on proton production by electrochemical reactions involving water molecules adsorbed at the organic/oxide interface, followed by proton diffusion through the oxide.[9,10]

We have considered $Si^+/SiO_2$/Pentacene/Au junctions that are the two terminal pendants of thin film transistors with the same layer structure. We study the dynamic electrical response of these junctions under ambient conditions over a large frequency window as function of a superimposed dc voltage ($V_{dc}$). This technique has proven in the past to be powerful because of its sensitivity and its ability to separate the different process involved.[11,12] It was already applied to several organic based junctions[13,14] but the results were interpreted based on models relying on the Shokley-read-Hall statistics not appropriated to organic materials.[11,12] Compared to these previous works[13,14] our frequency range is more extended, going down to $10^{-1}$ Hz, and the analysis of our data insist on the crucial role played by the interactions of the defects with their surroundings which determine the shape of the measured response functions.[15-18] We have identified two types of defects with different dynamic properties: one located inside the oxide shows a Debye type of response, the other located at the oxide/organic interface, or in the bulk of the organic, shows a Cole-Cole type of response.[17,19,20] At lower frequencies (< 100 Hz) we observe anomalous Low Frequency Dispersion (LFD)[21] that is consistent with the ionic diffusion current proposed in Refs. [9,10] but instead of normal diffusion considered in these works, our results evidence fractional diffusion to occur.[22,23]

## 2. Material and methods.

### 2.1. *Device preparation.*

The samples are fabricated using 70 (± 7) nm thick $SiO_2$ layer which was thermally grown on a heavy phosphorus doped n-typed silicon (100) (0.001 – 0.003 $\Omega$ cm). Prior to the pentacene deposition, the substrate was cleaned in an ultrasonic bath with acetone and isopropanol, and then blown with dry nitrogen. Subsequently, it was exposed to UV-ozone for 20 min. Then the thermal evaporation of the 40 nm thick pentacene organic active layer was done under the base pressure of $2.8 – 3.10^{-6}$ mbar. The deposition rate of pentacene was maintained at 0.1 Å/s on the substrate kept at room temperature. After the deposition of the pentacene, the sample was transferred to the metallization chamber in order to deposit 120 nm of gold electrodes.

### 2.2. *Admittance measurements.*

We have considered about 20 devices in ambient conditions showing all the same qualitative features. We present below results mostly for one of then that is very representative. For each sample a small part of the surface is left free of pentacene. This allows us to access on the $Si^+/SiO_2/Au$ junctions that constitute the "reference" measurements which can be compared with the organic based systems. We have measured the admittance in the frequency range $10^{-1} – 10^5$ Hz as function of a static potential ($V_{dc}$) varying between -20 V and +20 V. The $V_{dc}$ bias and the small superimposed $V_{ac}$ signal (100 mV) was applied to the gold electrode. Three different sizes of $9.10^{-4}$, $3.6.10^{-3}$ and $10^{-2}$ cm$^2$ of contact electrode were accessible on our devices but no size effects were noticed. The complex admittance, $Y(\omega)$, was measured using a frequency response analyzer (Solartron 1260) coupled with an interface dielectric (Solartron 1296) that provides directly the parallel conductance and capacitance. We write the response as[16]

$$\frac{Y(\omega)}{\omega} = j(C'(\omega) - jC''(\omega)) = j(\chi'(\omega) + C_\infty) + \chi''(\omega) \qquad (1)$$

where all the terms are supposed to be normalized by the contact surface area. $C'$ contains information about the polarization properties, $C''$ about the energy dissipated by the system during the polarization process. $\chi$ is the part of the dielectric susceptibility that includes the response of the slow species (slow ions, permanent dipoles,…). $C_\infty$ is the part of the capacitance that includes the response of the fast species that follow instantaneously the electric field at the applied frequencies (electrons, polarons, phonons,…).

## 3. Results and discussions.

Typical results are shown in Fig. 1a and 1b giving $C'$ and $C''$, respectively, for different dc voltages ($V_{dc}$). The results for the corresponding reference junction are also shown on the same graphs. These results show a loss part (Fig. 1b) structured and weakly frequency dependent with a minimum loss $\sim 10^{-10}$ S.s.cm$^{-2}$ (tan$\delta$ about $10^{-3}$).[20] As for all the junctions that we have investigated, our data clearly show that the admittance is decomposed in two qualitatively different types of independent contributions (see Figs. 1). The first type of contributions ($\chi_{Ion}$) observed at low frequencies (<100 Hz) for all of our devices shows a decrease both in the real and the imaginary part of $Y(\omega)/\omega$ following a fractional power law. We identify this contribution with a LFD observed in systems where the polarization is controlled by currents of slow charges.[21] The second type of contribution ($10^2$-$10^5$ Hz) is dipolar type that manifests itself as a peak in the loss part of the admittance.[16,17,20] Our data are very well reproduced if we consider only two different dipolar contributions ($\chi_{Organic}, \chi_{SiO_2}$). Because they are visible for different range of $V_{dc}$ they can be attributed to defects located one in the bulk of the $SiO_2$ and the other on the organic/oxide interface or within the organic film. The dielectric susceptibility of Eq. (1) is then written as a sum of three components

$$\chi(\omega) = \chi_{SiO_2}(\omega) + \chi_{organic}(\omega) + \chi_{Ion}(\omega) \tag{2}$$

The relative amplitude of each of these components fluctuates from device to device and varies with $V_{dc}$. We do the same analysis for the "reference" junctions for which no contribution of the pentacene appears.

The three contributions of Eq. (2) are detailed farther. With the Eqs. (1) and (2), we accurately fit the complex admittance at every $V_{dc}$ as will be seen below. In particular, the fitting procedure gives us the $C_\infty$-$V_{dc}$ curves that we first comment before presenting the different contributions of Eq. (2).

3.1. $C_\infty$-$V_{dc}$ characteristics.

An example of $C_\infty$-$V_{dc}$ curves that corresponds to the measurements of Figs. 1 is shown in Fig. 2. This quantity contains only the contributions of the fast degrees of freedom (hole, polarons,…) to the polarization and gives information about the electrostatic properties of a pure junction where all the slow contributions contained in $\chi$ have been removed.

The infinite frequency capacitance of the "reference" junctions has been shown to be bias independent, as it should be, and shows good agreements with the theoretical expectations: the case of Figure 1a, for instance, corresponds to an oxide of 77 nm width. The $C_\infty$-$V_{dc}$ curves of the pentacene based junctions (Fig. 2) appear to be similar to the ones of classical

Metal/Oxide/Semiconductor junctions but we have to keep in mind that the organic semiconductors differ noticeably from their inorganic counter part.[24,25] First, the organic semiconductor is not doped: so, charges come from the electrodes and by changing the applied voltage we modify both the charge density and the charge profile within the organic film. Moreover, the organic semiconductors are strongly polarizable and the free charge carriers are polarons, or bipolarons, that can not be described as easily as the quasi-electrons or quasi-holes of Si.[24,25] Nevertheless, we verify that for sufficiently strong $V_{dc}$, enough charges have been accumulated at the pentacene/oxide interface up to the full accumulation regime. The capacitance of the junction is so given by the oxide capacitance only. In decreasing $V_{dc}$ the carrier density decreases (Fig 2) and drives continuously the system from the accumulation to the full depletion (starting from $V_{dc}$=-5 V). Finally for small enough potential values the organic layer is well free of charges and reacts as a dielectric. In this regime the equivalent capacitance corresponds to the series of oxide and pentacene capacitances which is lead to a pentacene dielectric constant of about 3.5, in agreement with values recently reported in literature.[26] It should be stressed that, as commonly observed for organic field effect transistors, our junctions are not ambipolar (p-semiconductor). Even after applying a strong negative potential ($V_{dc}$<-40 V) the capacitor remains in the depletion regime: no negative charges are accumulated at the interface. A last remark concerns the shape of the $C_{\infty}$-$V_{dc}$ curve: our data shows a transition region between depletion and accumulation much smoother than the one commonly observed in Si based MOS junctions that could be due to the differences between the free charges found in these two systems.[11,12]

3.2. *Dipolar contributions*.

The two first contributions of the dielectric susceptibility (Eq. (2)) are of dipolar type. In general, the term dipolar should be understood in a broad sense referring to the frequency range where the responses take place. They are associated to particular entities that could be permanent dipoles carried by impurities present in the structure or traps that can be successively charged and discharged, for instance. In our cases, since these two contributions are also seen in the complete depletion regime where no free charges are present in the structure, they should be attributed to permanent dipoles. These dipoles may be seen as harmonic oscillators that oscillate under the action of $V_{ac}$ giving contributions to the polarization.[17,18] Moreover, they are not isolated systems but instead interact with their surroundings. These interactions cause energy dissipation and the characteristics of the surroundings (the bath) influence the dynamical properties of the oscillators.[17,18] Roughly speaking, if the bath has the faculty to forget instantaneously the perturbations induced by the oscillators (compared to the characteristic time of observation),[27] the dielectric susceptibility will be of Debye type. On the contrary, if the bath possesses a long term memory of these perturbations,[27] the time response of the oscillators will be slow down that shows up in the dielectric susceptibility as fractional power laws. This is a Jonscher type of response [15] that could be modeled by several empirical functions.[19,28,29] In our junctions the two types of response are observed depending on the location of the oscillators.

A single dipolar contribution is seen in the admittance of the "reference" junctions. It is of Debye type

$$\chi_{SiO_2}(\omega) = A_{SiO_2}\left(1 - j\omega\tau_{SiO_2}\right)^{-1} \tag{3}$$

$A_{SiO_2}$ is the amplitude, proportional to the density of such dipoles.[17] $\tau_{SiO_2}$ is the characteristic relaxation time of these dipoles that is found between $10^{-5}$s and $10^{-7}$s depending on cases. Two examples that differ by their characteristic relaxation time are shown: a first one in Figs. 1 with $\tau_{SiO_2} \approx 10^{-6}$ s, and a second one that presents also fewer fluctuations is presented in Figs. 3 with $\tau_{SiO_2} \approx 10^{-5}$ s. As a remark we may notice that the loss part of the admittance, $C''$, of the two examples differ by one order of magnitude: in case of Figs. 1 it is close to the sensitivity limitation of our experiment what explains the wide fluctuations observed in the data. In the case of the organic MOS junctions, in the accumulation regime, again the bulk of the oxide only is probed and indeed, we obtain the same response (Eq. (3)), with the same characteristic frequency (see Figs. 4a and 4b). The two parameters, $A_{SiO_2}$ and $\tau_{SiO_2}$, are potential independent with fixed values at all $V_{dc}$: $A_{SiO_2} \approx 7.10^{-9}$ and $1/\tau_{SiO_2} \approx 2.10^6$ Hz for the example shown in Figs. 1 and 4. As a remark, we may notice that the same kind of response was already obtained in the past for MOS capacitors that was interpreted in terms of charging and discharging of surface states.[30] Since the reference junctions are of metal/oxide/metal type, the response should concern different impurities, as already mentioned, located deep in the bulk and then not accessible by tunneling.

As usual in MOS capacitor, the organic/oxide interface is probed when the junction is driven from accumulation to depletion regime. Decreasing progressively the voltage from accumulation, a second dipolar contribution appears at lower frequencies (see Figs. 1 from $10^2$ to $10^4$ Hz). It is of Jonscher type and well described by the Cole-Cole susceptibility

$$\chi_{Organic}(\omega) = A_{Organic}\left(1+\left(-j\omega\tau_{Organic}\right)^{\alpha}\right)^{-1} \qquad (4)$$

where $0 < \alpha < 1$.[19] The amplitude, $A_{Organic}$, the characteristic frequency, $1/\tau_{Organic}$, and $\alpha$ ($\approx$ 0.5) change slightly with $V_{dc}$. We discuss in the following only the changes of characteristic frequencies from which we are able to obtain some information.

The two dipolar contributions may be modeled by damped harmonic oscillators in interaction with fluctuating surroundings.[17,18] It was shown in Ref. [17] that the characteristic frequency of each oscillator depends on the electric field in the following way

$$1/\tau_{SiO_2/Organic}(\omega) \propto \vec{\mu}_{SiO_2/Organic}\vec{E}_{local} \qquad (5)$$

$\vec{\mu}_{SiO_2/Organic}$ is the permanent dipole carried by the corresponding impurity. $\vec{E}_{local}$ is the electric field seen by the dipole considered. We have to include in Eq. (5) the local electric field instead of the applied field as done in Ref. [17], because the polarons may screen the applied potential. According to Eq. (5), the behaviors of the characteristic frequencies as function of $V_{dc}$ give information about the dipole orientation and the electrostatic properties of the junctions. $1/\tau_{SiO_2}$ does not vary with $V_{dc}$. Since the data correspond to the sum of the responses of many impurities of the same kind, it means that there is no preferential orientation for the permanent dipoles carried by these defects located in the oxide. On the contrary, $1/\tau_{Organic}$ varies as shown in Fig. 5. We may conclude that the projection of the corresponding permanent dipoles is directed toward the substrate. Moreover, we clearly see different behaviors between the depletion regime ($V_{dc} < -5$ V), where the variations are linear, and the transition to the accumulation regime ($V_{dc} > -5$ V), where non-linearities are evidenced. Notice that the figure 5 is limited to the negative values of the dc potential for simplicity because this is the region where the Cole-Cole contribution is more pronounced. The nature of the relaxing quantities responsible for these two contributions remains to be clarified. As examples, it could be due to substitution atoms that induced permanent dipoles, for the Debye contribution, and to water molecules adsorbed at the oxide/organic interface, for the Cole-Cole contribution.

3.3. *Ionic contributions*.

The last contribution of the dielectric susceptibility (Eq. (2)) is a LFD type of contributions which is usually attributed to diffusion current of slow charges.[21] Moreover, in some heterostuctures it was shown to be related to adsorbed water layer on one of the interfaces in the inside of the structure.[21] Under ambient condition such layer is expected to be formed in our systems at the surface oxide: water molecules is known to diffuse through the organic film.[9] They would then react with the Si-OH bonds present at the surface. This layer is supposed to be the source of electrochemical reactions such as the one proposed in Refs. [9,10] that could be a possible origin to the LFD as explained in details below. When hole polarons (h) – or simply hole in the case of the reference junctions - are accumulated at this interface, protons are produced according to the coupled reactions $2H_2O + 4h \rightarrow 4H^+ + O_2(g)$ and $2H^+ \rightarrow 2h + H_2(g)$. The protons $H^+$ then migrate through the oxide, probably in the $H_3O^+$ state. The diffusion of particles through amorphous systems is known to follow, in general cases, a continuous time random walk in which the ions may remain after each jump at the same position for infinitely long time.[31,23] This type of random walk can be described by the fractional diffusion equation[23,32] characterized by a parameter $\beta$ ($0 \leq \beta \leq 1$), the case $\beta = 1$ corresponding to normal diffusion considered in Refs. [9,10]. This parameter controls the dynamics of the ions and can be measured in our experiment.

Because of the diffusive intrusion of protons in the oxide the capacitance becomes a function of time, $C(t)$. The time needed for an ion to cross the oxide layer can be estimated to $10^7$ s, by taking $K_1 \approx 2.10^{-19}$ cm$^2$s$^{-1}$ given in Ref. [10], but is certainly longer for anomalous diffusion. As a consequence, during the time of our experiment, the protons remain very close to the interface and we may reasonably assume that the changes of capacitance are a linear function of the amount of ions inside the oxide. Taking the time derivative, we may then write (see appendix)

$$\dot{C}(t) = -A_C J_\beta(0,t) \qquad (6)$$

$J_\beta$ is the anomalous diffusion current of protons taken at the interface ($x = 0$) and the constant $A_C$ takes account for the effects of the ions on the capacitance. According to Eq. (6), the dielectric susceptibility of the LFD is directly related to the admittance of the ionic anomalous diffusion, $Y_{Ion}$, which was calculated in Refs. [22,33]. In the limit of high frequencies (compared to the characteristic frequency $10^{-7}$ Hz for ion diffusion) that corresponds to our experiments, we finally obtain:

$$\chi_{Ion}(\omega) = -\frac{A_c}{j\omega} Y_{Ion}(\omega) \approx A_{Ion}(-j\omega)^{-\beta/2} \qquad (7)$$

that reproduces very well our data (Figs 3 and 4). The detailed expression of the amplitude $A_{Ion}$ can be found in the appendix. For the reference junctions we get $\beta = 1$ that corresponds to normal diffusion (see Figs. 3). For the organic based junctions we find anomalous diffusion with $\beta$ between 0.2 and 0.4. In the example of Fig. 4, the Cole-Cole contribution is strong enough to spoil the characteristics of the LFD contribution that are more apparent in some other examples such as the one shown in Fig. 6.a. We find $\beta = 0.4$ in the example of Figs. 4 and $\beta = 0.3$ in the example of Fig. 6.a. Moreover, we notice that the amplitude, $A_{Ion}$, increases when $V_{dc}$ decreases in a way shown in Fig. 6.b that corresponds to the junction investigated in Figs. 1 and 4. This behavior is in agreement with Eq. (A12) of the appendix that shows that $\chi_{Ion}$ is inversely proportional to the proton density at the interface. According to the chemical reaction described above this density is directly related to the density of holes accumulated at the interface – a linear relation is assumed in Ref. [10]. The fact that the diffusion characteristics are different in the two types of junctions may be explained by the nature of the interfaces.

## 4. Conclusions.

In summary, we have measured the low frequency admittance of Metal/Oxide/Organic semiconductor junctions and identified both dipolar and diffusive current contributions to the polarization. The dynamical responses of the polarizable species (dipoles or ions) are strongly influenced by their surroundings. If they are markovian (without memory) we get a Debye response for the dipoles and a normal diffusion for the ions as measured for "reference" junctions. If they are non-markovian (with memory) the responses are more complex; in their simplest forms we get a Cole-Cole response for the dipoles and a fractional diffusion for the ions. Our data show anomalous behavior in presence of organic layer. We believe these observations could be the signature of particular structural organizations which can be located either, in the organic layer itself or, in the organic/oxide interface.

*Acknowledgement*
We thank Nicolas Clement for fruitful discussion and the *Region Champagne-Ardenne* for financially supporting of PhD of Romuald Ledru. This work has been supported by the ANR under contract N°ANR-09-BLAN-0329-02 (CADISCOM).

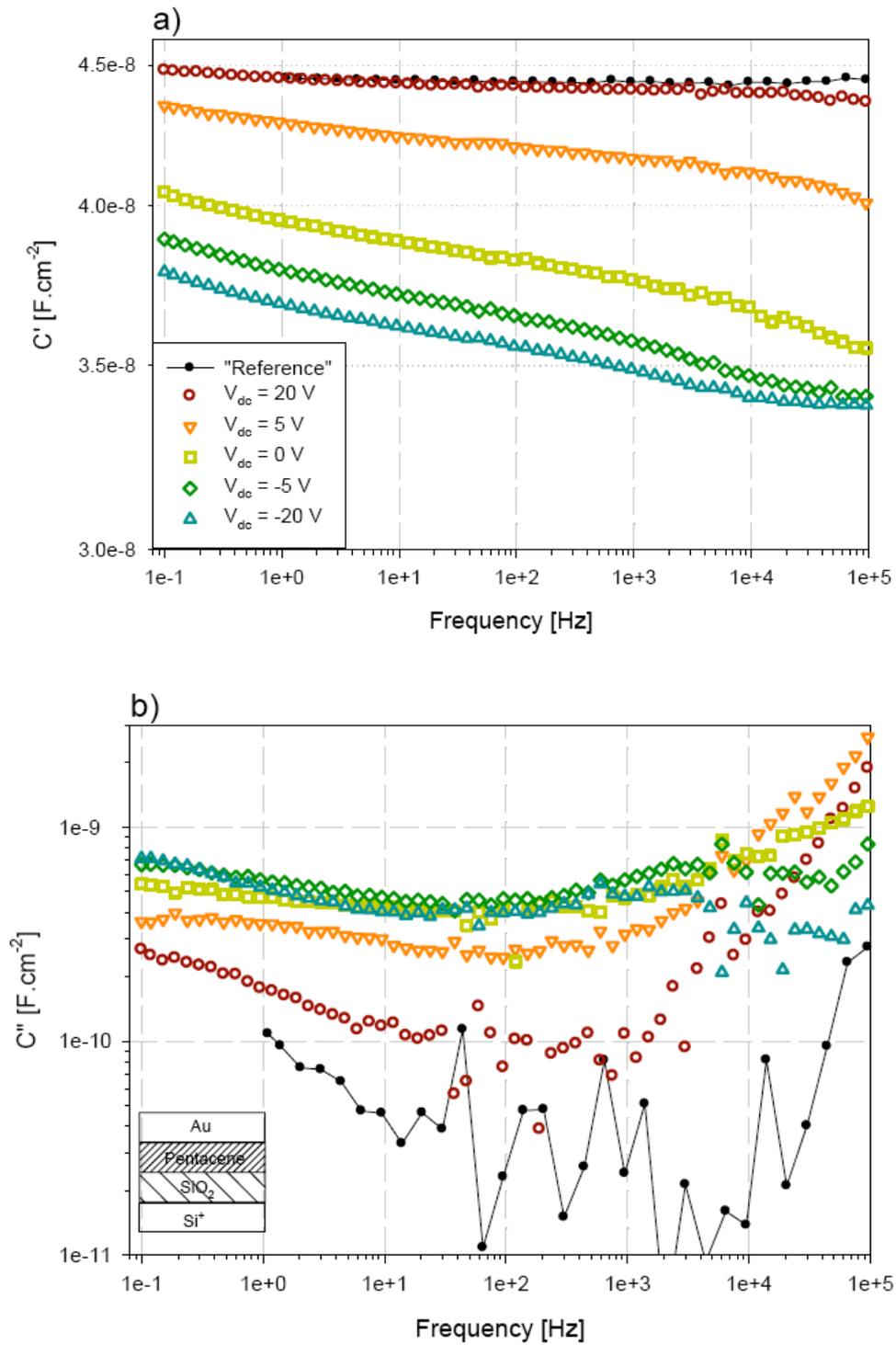

***Figure 1.*** *Typical capacitance, a) real part, C′, and b) imaginary part, C″, for $V_{dc}$=20 V to -20 V. Results for the reference junction at $V_{dc}$ = 0V are also shown (dots). Insert of b): Schematic of the layer structure of the junctions.*

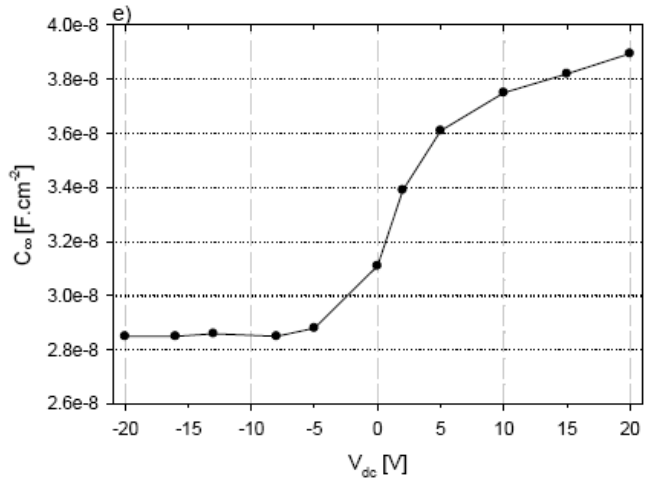

**Figure 2.** Infinite frequency capacitance, $C_\infty$, extracted from the results shown in Fig. 1, as function of $V_{dc}$

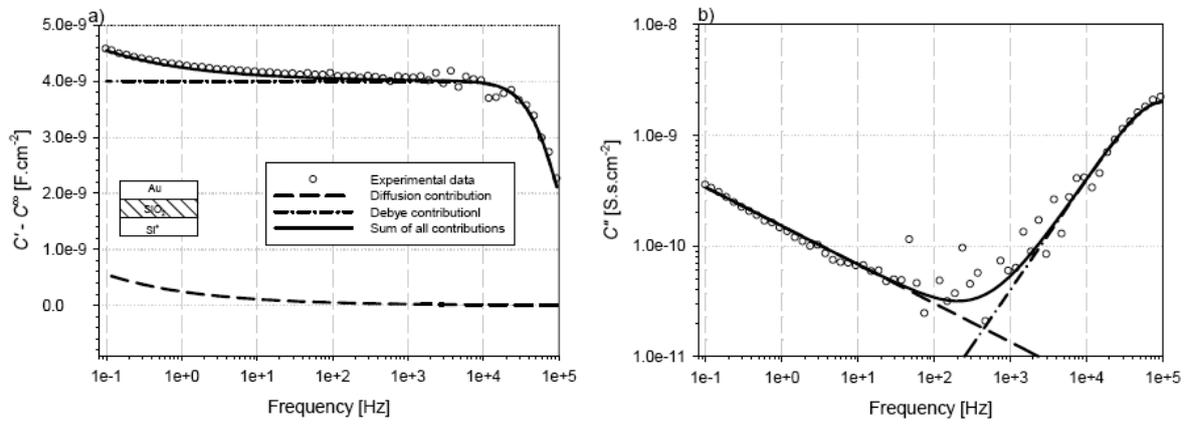

**Figure 3.** Typical capacitance of a "Reference" $Si^+/SiO_2/Au$ a) real part, b) imaginary part at $V_{dc}=0$ V. The experimental data are fitted with two contributions: a Debye contribution for $SiO_2$ defects (Eq. (3)) and a LFD contribution for ion diffusion in the oxide (Eq. (6)). The fitting parameters are $A_{SiO_2}=4.10^{-9}$, $\tau_{SiO_2}=10^5 s$, $A_{Ion}=5.5\ 10^{-10}$ and $\beta=1$.

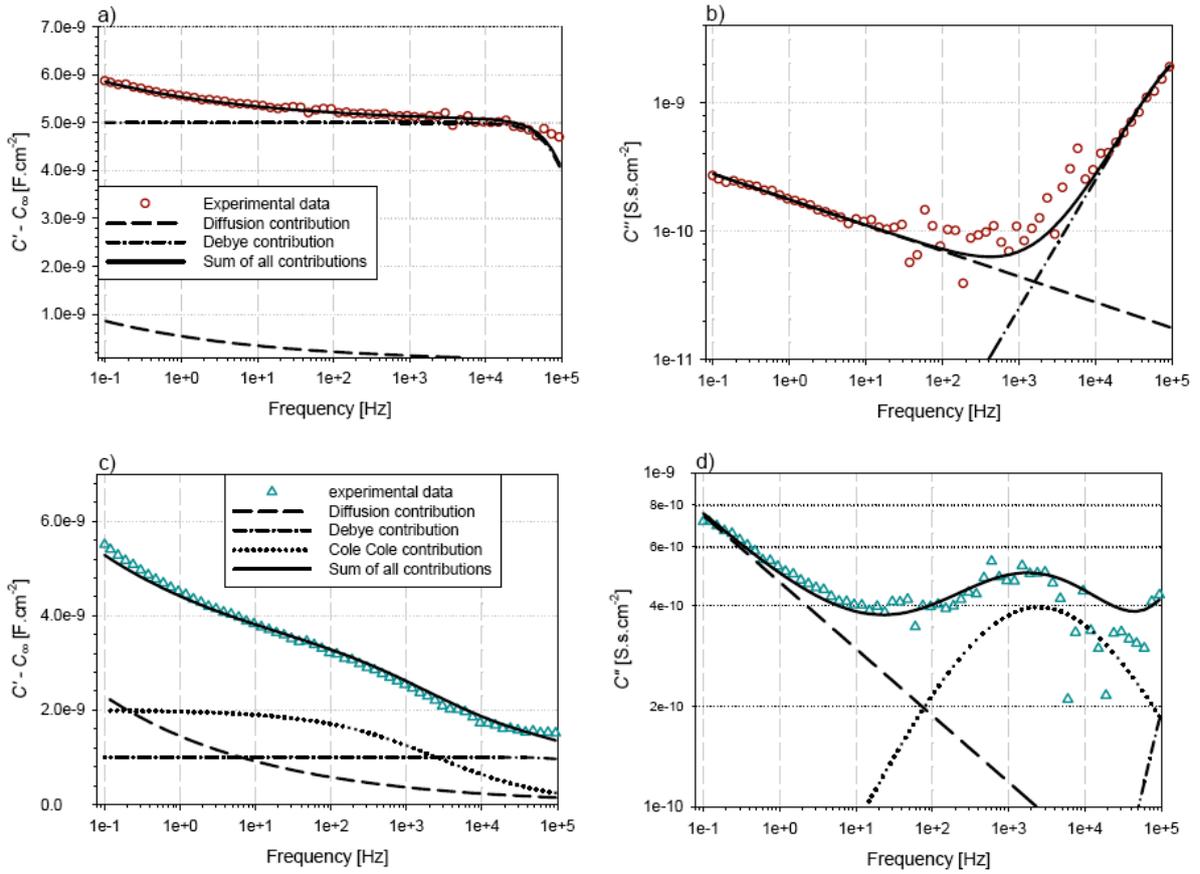

***Figure 4.*** *Focus on two particular examples of Fig. 1 at $V_{dc}=+20\ V$ a) real part, b) imaginary part, and at $V_{dc}=-20\ V$ c) real part, d) imaginary part. The experimental data are fitted with two or three contributions: a Debye contribution for $SiO_2$ defects (Eq. 3) - doted-dashed lines, a Cole-Cole contribution for organic defects (Eq. 4) – doted lines, and a LFD contribution for ions at the $SiO_2$/pentacene interface (Eq. 6) – doted line. The values of the parameters are discussed in the text.*

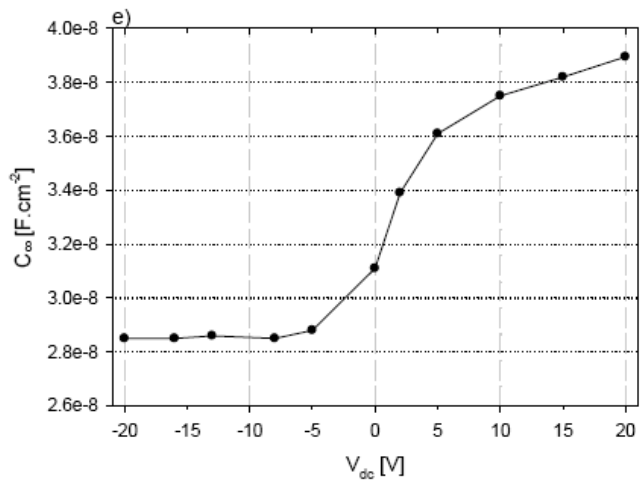

***Figure 5.*** *$1/\tau_{organic}$ as function of $V_{dc}$: a change of behavior may be noticed between the depletion regime ($V_{dc} < -5\ V$) and the transition to the accumulation regime ($V_{dc} > -5\ V$).*

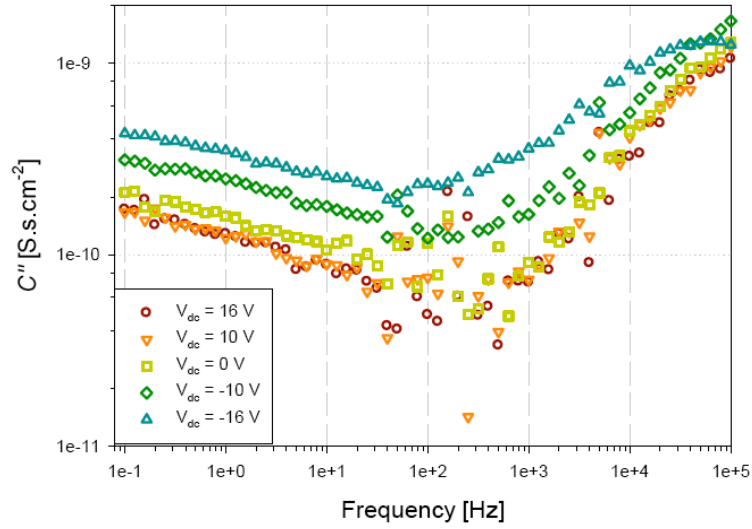

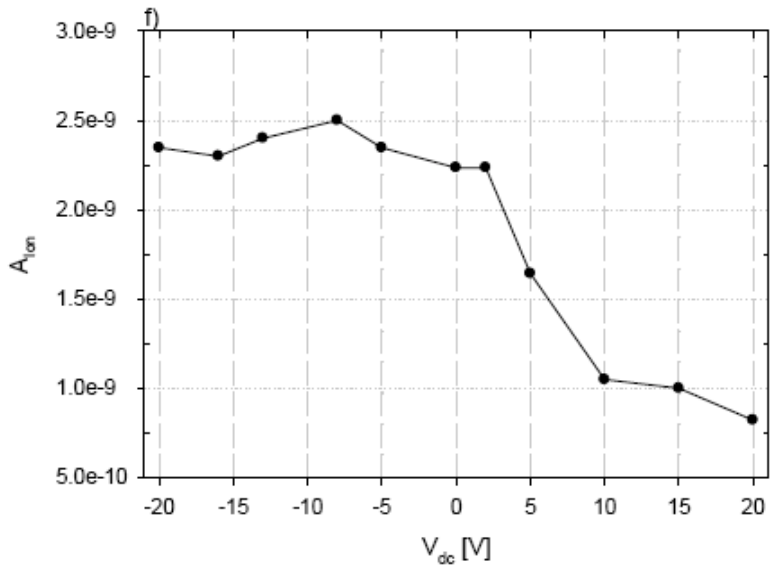

*Figure 6.* a) C″ for a junction with a weak Cole-Cole contribution showing more clearly the LFD contribution at all $V_{dc}$: Some examples are shown from the accumulation to the depletion regime. The β parameter is bias independent - in this case β = 0.3, and the amplitude of the ionic contribution increases when $V_{dc}$ decreases. b). Amplitude of the ionic contribution, $A_{Ion}$, as function of $V_{dc}$ for the example shown in Figs. 1 and 4.

**Appendix: Model for the anomalous Low Frequency Dispersion (LFD).**

We note $p(x,t)$ the time dependant volume density of protons in the oxide at distance $x$ from the organic/oxide interface. The thickness of the oxide layer is $L \approx 70$nm. As it is usual for amorphous systems, the protons are assumed to follow a continuous time random walk through the oxide.[31,23] It means that after each jump the protons may remain at the same position for infinitely long time. In practice this could happen in deep traps. This type of random walk can be described by the fractional diffusion equation [32]

$$\frac{\partial p(x,t)}{\partial t} = {}_0D_t^{1-\beta} K_\beta \frac{\partial^2 p(x,t)}{\partial x^2} = \frac{K_\beta}{\Gamma(\beta)} \frac{\partial}{\partial t} \int_0^t dt' \frac{1}{(t-t')^{1-\beta}} \frac{\partial^2 p(x,t')}{\partial x^2} \tag{A1}$$

${}_0D_t^{1-\beta}$ is the fractional Riemann-Liouville operator with definition given by the second equality of Eq. (A1), $\beta$ is a constant such that $0 \leq \beta \leq 1$, $\Gamma$ is the Gamma function and $K_\beta$ generalizes the diffusion coefficient with dimension cm$^2$.s$^{-\beta}$. The case $\beta = 1$ corresponds to normal diffusion. This density should also fulfill the charge conservation equation

$$\frac{\partial p(x,t)}{\partial t} = -\frac{\partial J_\beta(x,t)}{\partial x} \tag{A2}$$

$J_\beta$ is the anomalous diffusion current of protons that may be written as [34]

$$J_\beta(x,t) = -q \frac{K_\beta}{\Gamma(\beta)} \frac{\partial}{\partial x}\left[ t^{\beta-1} p(x,0) + \int_0^t dt' p(x,t')(t-t')^{\beta-1} \right] \tag{A3}$$

$q$ is the electric charge of the diffusing ions. On the contrary to normal diffusion ($\beta=1$), the current is non-local in time: at each time it depends on the previous history. Moreover, the first term of Eq. (A3) shows that the initial condition contributes at all times.

Because the protons diffuse through the oxide, the capacitance of our junction becomes time dependant, $C(t)$. The proton diffusion is very slow: it takes about $10^7$ s for an ion to cross the oxide so that during the time of our experiments, they all remain close to the organic/oxide interface.[10] We therefore suppose

$$C(t) = C(0) - A_C \int_0^L dx\, p(x,t) \tag{A4}$$

The ions modify the capacitance by Coulomb interaction but since the ions are all approximately at the same distance from the interface the complicated space dependence due to these interactions may be reduced to a single constant $A_C$. Taking the time derivative of Eq. (A4) and with the help of the charge conservation equation (Eq. (A2)) we obtain

$$\dot{C}(t) = -A_C J_\beta(0,t) \tag{A5}$$

The dot is for the time derivative. We have considered that $J_\beta(L,t) = 0$ since all the ions remain near $x = 0$.

The ions continuously diffuse through the oxide. Applying in addition a small ac potential introduces a perturbation to the main contribution. We write

$$p(x,t) = p_0(x,t) + p_1(x,t) \tag{A6}$$

$p_0$ is the dc component, $p_1$ the ac perturbation. In the same way we can decompose the surface potential at the organic/oxide interface

$$\psi(t) = \psi_0(t) + \psi_1(t) \tag{A7}$$

To determine the admittance associated to the ion displacement we need to consider only the ac components; $p_0(x,t)$ and $\Psi_0(t)$ may indeed be considered as time independent in the time spend to do the experiments. Taking the Laplace transform of Eq. (A5) yields

$$\chi_{Ion}(s) = -\frac{A_C}{s} Y_{Ion}(s) \tag{A8}$$

where the admittance of the ionic current is defined as

$$Y_{Ion}(s) = \frac{\tilde{J}_\beta(0,s)}{\tilde{\psi}_1(s)} \tag{A9}$$

$\tilde{J}_\beta(0,s)$ and $\tilde{\psi}_1$ are the Laplace transforms of the diffusion current, $J_\beta$, at $x = 0$ and of $\Psi_1$, respectively. Eq. (A9) was solved in Ref. [22] assuming a linear relation between $p_1(x,t)$ at the interface and $\psi_1(t)$. With ideal conditions for the electrochemical reaction it may be written[33]

$$\psi_1(t) = \frac{k_B T}{p_0(0,0)} p_1(0,t) \tag{A10}$$

$T$ is the temperature and $k_B$ the Boltzmann constant. Different boundary conditions at the other interface were considered by the authors of Ref. [22] but, it is important to stress that at the frequencies we consider in this work they do not play any role. We quote their solution for absorbing boundary conditions, $p_1(L,t)=0$. In terms of frequency instead of Laplace coordinate we finally get

$$\chi_{Ion}(\omega) = -\frac{A_c}{j\omega} Y_{Ion}(\omega) = \frac{qk_B T K_\beta}{L} \frac{\omega_d^{1-\beta}}{p_0(0,0)} \left(\frac{-j\omega}{\omega_d}\right)^{-\beta/2} \frac{1}{\tanh\left((-j\omega/\omega_d)^{\beta/2}\right)} \qquad (A11)$$

with $\omega_d = (K_\beta / L^2)^{1/\beta}$. At high frequencies, $\omega/\omega_d \to +\infty$, this equation gives

$$\chi_{Ion}(\omega) \approx q \frac{k_B T K_\beta}{L} \frac{\omega_d^{1-\beta}}{p_0(0,0)} \left(\frac{-j\omega}{\omega_d}\right)^{-\beta/2} = A_{Ion}(-j\omega)^{-\beta/2} \qquad (A12)$$

that corresponds to Eq. (7) of the main text. It would be the same with open boundary conditions, for instance.